\documentclass[aps,twocolumn,amsmath,amssymb,prx,showpacs]{revtex4-1}

\usepackage{graphicx}
\usepackage{comment}
\usepackage[normalem]{ulem}

\makeatletter
\def\blfootnote{\xdef\@thefnmark{}\@footnotetext}
\makeatother

\begin{document}
%
%
\title{Bulk Rotational Symmetry Breaking in Kondo Insulator SmB$_6$}
\author{Z. Xiang$^{1}$, B. Lawson$^1$, T. Asaba$^1$, C. Tinsman$^1$, Lu Chen$^1$, C. Shang$^2$, X. H. Chen$^2$, Lu Li$^{1*}$}
\affiliation{
$^1$Department of Physics, University of Michigan, Ann Arbor, Michigan 48109, USA\\
$^2$Hefei National Laboratory for Physical Sciences at Microscale and Department of Physics, University of Science and Technology of China, Hefei, Anhui 230026, China and Key Laboratory of Strongly-coupled Quantum Matter Physics, Chinese Academy of Sciences, Hefei, Anhui 230026, China}

\date{\today}
\begin{abstract}
Kondo insulator samarium hexaboride (SmB$_6$) has been intensely studied in recent years as a potential candidate of a strongly correlated topological insulator. One of the most exciting phenomena observed in SmB$_6$ is the clear quantum oscillations appearing in magnetic torque at a low temperature despite the insulating behavior in resistance. These quantum oscillations show multiple frequencies and varied effective masses. The origin of quantum oscillation is, however, still under debate with evidence of both two-dimensional Fermi surfaces and three-dimensional Fermi surfaces. Here, we carry out angle-resolved torque magnetometry measurements in a magnetic field up to 45 T and a temperature range down to 40 mK. With the magnetic field rotated in the (010) plane, the quantum oscillation frequency of the strongest oscillation branch shows a four-fold rotational symmetry. However, in the angular dependence of the amplitude of the same branch, this four-fold symmetry is broken and, instead, a twofold symmetry shows up, which is consistent with the prediction of a two-dimensional Lifshitz-Kosevich model. No deviation of Lifshitz-Kosevich behavior is observed down to 40 mK. Our results suggest the existence of multiple light-mass surface states in SmB$_6$, with their mobility significantly depending on the surface disorder level.
\end{abstract}

\blfootnote{Corresponding authors: L. Li (luli@umich.edu)}


\maketitle                   

In Kondo insulators, the physics is controlled by the strong many-body interactions \cite{riseborough2000heavy}. The hybridization between the localized \textit{f} electrons and conduction \textit{d} electrons causes the formation of Kondo singlets, which leads to a quench of local-magnetic-moment characteristics. Also, a narrow hybridization gap is developed at low temperature, resulting in a crossover from metallic to insulating behavior. In recent years, topological nontriviality is suggested to be hosted by Kondo insulators \cite{dzero2010topological,dzero2012theory}. The opposite parity in the \textit{f} band (odd) and \textit{d} band (even) protects a band inversion similar to that in normal $Z_2$ topological insulators. In particular, the very large spin-orbit coupling in the renormalized \textit{f} electrons can give a system ground state with ``nontrivial" topological order, i.e., a different topological invariant from that in vacuum. As a result, a gapless two-dimensional (2D) Dirac electron state, known as the topological surface state, has to exist at certain high-symmetry points in the surface Brillouin zone. Such predictions point the Kondo insulators out as promising candidates of interaction-driven topological insulators, subsequently make this family a focus of attention in condensed matter physics.

\begin{figure*}[htbp!]
\centering
\includegraphics[width=0.75\textwidth]{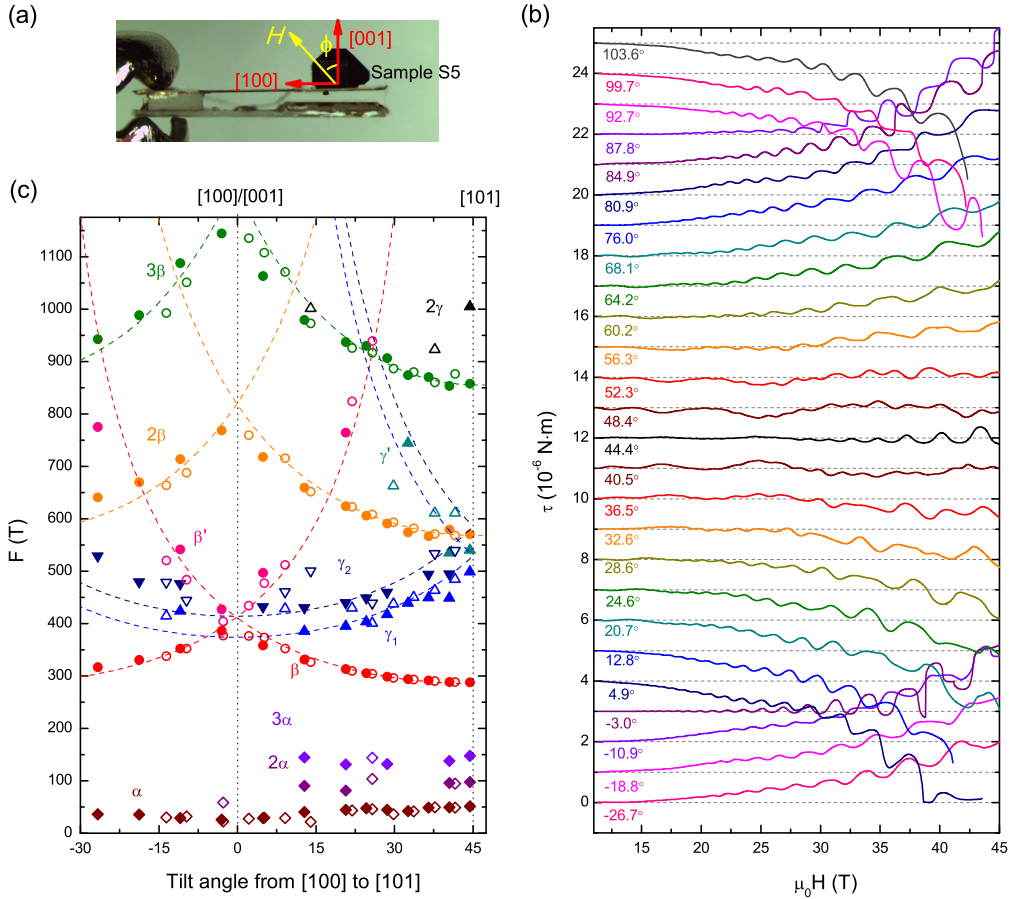}
\caption{Torque measurements on SmB$_6$. (a) Photograph showing the beryllium-copper cantilever and the SmB$_6$ single crystal S5 we used for the torque magnetometry measurement. In this setup, the magnetic field is rotated in the (010) plane of the sample. Arrows sketch the definition of the tilt angle $\phi$. (b) Magnetic torque of sample S5 measured up to 45 T at different tilt angles. The torque curves are shifted vertically for clarity. The absolute value of torque is calibrated by the sample weight. (c) Angular dependence of all resolved dHvA oscillation frequencies on the fast Fourier transform (FFT) of $M_{eff}$ with field rotated in (010) plane. FFT peaks are indexed with the same labels as in Ref. \cite{li2014two}. Solid symbols denote the data points at $\phi < 45^\circ$, while hollow symbols are data taken at $\phi > 45^\circ$. Harmonics of branches $\alpha$, $\beta$ and $\gamma$ are presented by diamonds, circles and triangles, respectively. Dashed lines are fittings based on 2D FS model: $F$ = $F_0$/$\cos(\phi-\phi_0)$. For branch $\beta$, $\phi_0$ = $\pm$45$^\circ$ and $F_0$ = 285 T. The two ``split" branches $\gamma_1$ (blue) and $\gamma_2$ (navy) both have symmetric axis along [100], i.e., $\phi_0$ = 0$^\circ$ and 90$^\circ$, while $F_0$ is 374 T for $\gamma_1$ and 414 T for $\gamma_{2}$.
}
\label{FigTorque}
\end{figure*}

The cubic structured SmB$_6$, the very first confirmed member of Kondo insulators \cite{menth1969magnetic}, has been elaborately studied as the most feasible example of the electron-correlated three-dimensional (3D) strong topological insulator \cite{takimoto2011smb6,alexandrov2013cubic,lu2013correlated}. A large amount of experimental observations on this material have been published \cite{dzero2016topological}, with some giving hints of the topological surface state \cite{kim2014topological,thomas2016weak,xu2014direct}, though the decisive evidence is yet to be found. The most striking discovery in SmB$_6$ is the complicated de Haas-van Alphen (dHvA) oscillations detected at low temperature where the resistivity shows insulating behavior followed by a plateau below 3.5 K. While our group has reported quantum oscillations corresponding to 2D Fermi surface (FS) and light carriers that are consistent with the expectation on a typical topological surface state in aluminum-flux-grown samples \cite{li2014two}, another work based on the floating-zone(FZ)-grown sample claimed the oscillations have 3D characters thus bulk origin, and an abnormally enhanced quantum oscillation amplitude suggesting a deviation from the Lifshitz-Kosevich (LK) theory below $^3$He temperature \cite{tan2015unconventional}. To make it more confusing, no quantum oscillations have ever been observed in transport measurements \cite{li2014two,wolgast2015magnetotransport,chen2015magnetoresistance}. Several theories have been proposed to reconcile the puzzling experimental results \cite{alexandrov2015kondo,erten2016kondo,peters2016coexistence} as well as to explain the enriched exotic low-temperature behaviors in SmB$_6$~\cite{knolle2015quantum,zhang2016quantum,Pal,pixley2015global,baskaran2015majorana,erten2017skyrme}. The key problem to the confusion lies in the lack of a controlled study of the quantum oscillation amplitude.

In this work, we resolve the problem by mapping the oscillation amplitudes of DIFFERENT surfaces of the SAME crystal. We carried down magnetic torque measurements of flux-grown SmB$_6$ single crystals \cite{canfield1992growth} down to 40 mK in a rotating magnetic field up to 45 T.  Our result shows a broken rotating symmetry in the amplitude of the main dHvA oscillation branch, indicating a 2D nature of the electronic state. In addition, neither very high frequency oscillations with 3D behavior, nor abruptly enhanced dHvA amplitude suggesting failed a LK description is observed in any of our samples. These observations point to multiple 2D metallic states with small effective mass existing in Kondo insulator SmB$_6$.

By using the capacitive magnetic torque magnetometer shown in Fig. \ref{FigTorque}(a), we observe clear dHvA oscillations with the coexistence of different periods in our flux-grown SmB$_6$ samples (for details of sample preparation and experimental methods, see Appendix \ref{chap01}). The angle-resolved field dependencies of the magnetic torques $\tau$ are shown in Fig. \ref{FigTorque}(b). We convert the measured capacitance $C(B, \theta)$ to torque by the relation $\tau \propto 1/C$. The absolute value of the torque signal is calibrated by a zero-field rotation in which the weight of sample $m$ generates a change in the capacitance of $\Delta(1/C) \propto mgl\cos \phi$. Here $l$ is the length of the cantilever beam and $\phi$ is the sample tilt angle between the magnetic field and the crystalline [001] direction as described in Fig. \ref{FigTorque}(a).

There are several interesting features in the angle-dependent behavior of $\tau(B)$.  First, it is apparent that the non-oscillatory background of $\tau(B)$ changes its sign abruptly at $\phi$ = 0$^\circ$ and 90$^\circ$, while at $\phi$ = 45$^\circ$ there is another sign change but it is more smoother. We argue that this is most likely due to a bulk magnetic susceptibility anisotropy between the cubic [100] and [101] directions. Second, the dHvA oscillations on the torque curves also change sign at $\phi$ = 0$^\circ$, 45$^\circ$ and 90$^\circ$. This ``flipped" dHvA pattern across certain magnetic field directions (see $\tau(B)$ curves at $\phi$ = 40.5$^\circ$ and 48.4$^\circ$ in Fig. \ref{FigTorque}(b) as a typical reference) hardly reflects a sudden jump on the phase of quantum oscillation, while a more natural explanation is a direction change of the oscillatory torque vector $\vec{\tau}$ = $\vec{M} \times \vec{B}$ that happens at these angles. Both a 3D electronic system with susceptibility anisotropy along [100] and [101], and a 2D diamagnetic system, can exhibit such an oscillatory torque flip at $\phi = \pm(N/4)\pi$ ($N$ can be any integer). Third, the amplitude of dHvA oscillation is considerably large. At above 35 T, the oscillatory torque $\Delta \tau$ is roughly (0.5-1)$\times10^{-6}$ N$\cdot$m, corresponding to an effective magnetic moment $\Delta M_{eff}$ = $\Delta \tau$/$\mu_0 H$ of approximately (10-20)$\times10^{-9}$ A$\cdot$m$^2$.

Such a large dHvA oscillation amplitude basically rules out the possibility of ``false quantum oscillation" coming from the aluminum (Al) flux incorporated inside the sample (Appendix \ref{chap02}), but also gives some difficulties to the 2D surface state interpretation. In a standard analysis on 2D electron systems, the magnetization oscillation has an amplitude upper limit of $e\hbar n_{2D}/\pi m^*$ for a unit area. This dHvA amplitude is usually much larger than that in real materials since there are several damping factors needed to be taken into account \cite{shoenberg2009magnetic}. Here $n_{2D}$ is the 2D density of carriers that contribute to the dHvA oscillations. Using the electronic parameters calculated in our earlier work \cite{li2014two}, a 2D magnetic moment of $\sim$ 1$\times10^{-9}$ A$\cdot$m$^2$ is estimated for a surface area of 1 cm$^2$. Giving the millimeter size of our sample and the effective magnetic moment $\sim$ 1$\times10^{-8}$ A$\cdot$m$^2$, the discrepancy turns out to be roughly 2 order of magnitude. Moreover, the $M_{eff}$ discussed above is only the component perpendicular to the magnetic field, $M_{eff}$ = $M_{\perp}$, therefore smaller than the total magnetic moment. The puzzling large amplitude of the dHvA oscillations has been, however, reported in the confirmed 3D topological insulator Bi$_{1-x}$Sb$_x$ \cite{taskin2009quantum}, though the reason why the conventional estimation based on the surface carrier density failed there is unknown \cite{ando2013topological}. Further works looking into the peculiar magnetizing properties of topological surface state are needed to solve this question.

The fast Fourier transformation (FFT) of $M_{eff}$ show results that are consistent with our previous work \cite{li2014two}. Three main branches $F^\alpha$, $F^\beta$, $F^\gamma$ and their higher order harmonics are resolved. The FFT peak positions are plotted in Fig. \ref{FigTorque}(c) as a function of the angle between the applied field and the crystalline equivalent [100] directions in the cubic structure of SmB$_6$. For $F^\beta$ and $F^\gamma$, a fitting of $F = F_{0}/\cos(\phi - \phi_{0})$ can follow the behavior of $F(\phi)$ quite well, indicating a 2D nature of the related FSs. The value of $\phi_{0}$ hints that pockets $\beta$ and $\gamma$ have the symmetric axes along the equivalent [101] and [100] directions, respectively. A ``split" of peak $\gamma$ is observed in a wide angle range which may indicates a subtle magnetic breakdown (for details, see Appendix \ref{chap03}). These results are confirmed by repeated measurements in several samples and no sample dependence on the dHvA frequencies have ever been observed. Also, a recent tunneling spectroscopy study on SmB$_6$ single crystals reveals two Dirac-like surface bands on (100) surface and an additional one on (101) surface \cite{park2016topological}, which is in agreement with our observation of three bands. The smallest orbit $\alpha$, however, has indeterminate dimension and geometry (Appendix \ref{chap03}).

We do not resolve any dHvA frequencies higher than 2 kT (Appendix \ref{chap04}). The high-frequency components observed in FZ-grown SmB$_6$, which indicate large Fermi pockets with the size comparable to the area of Brillouin zone \cite{tan2015unconventional}, are confirmed to be absent in flux-grown samples. It should be pointed out that there are still some similarities between the quantum oscillation spectra in our flux-grown samples and those in FZ-grown crystals (see Appendix \ref{chap05} for details). In FZ-grown SmB$_6$, low frequency oscillations were also resolved and assigned to small orbits $\rho$ and $\rho'$ \cite{tan2015unconventional}. $\rho'$ shares the same angle range with our branch $\alpha$ and $\rho$ is close to $\beta$ and $\gamma$ in our FFT spectra. As mentioned in Ref. \cite{tan2015unconventional}, alternative possibilities are cylinder-like ``neck" sections in a 3D electronic structure, or elongated ellipsoidal FSs. Similar elaborated comparison of the angle dependence of $\beta$/$\gamma$ and $\rho$ has been made in Ref. \cite{erten2016kondo}, which shows inconclusive result in distinguishing the effectiveness of 2D and 3D model. Therefore, the angular dependence of the oscillation frequencies cannot determine the origin of quantum oscillations in SmB$_6$. As demonstrated later, the angular dependence of the oscillation amplitudes indicates that the oscillations most likely aries from the surface state.

\begin{figure}[htbp!]
\centering
\includegraphics[width=0.44\textwidth]{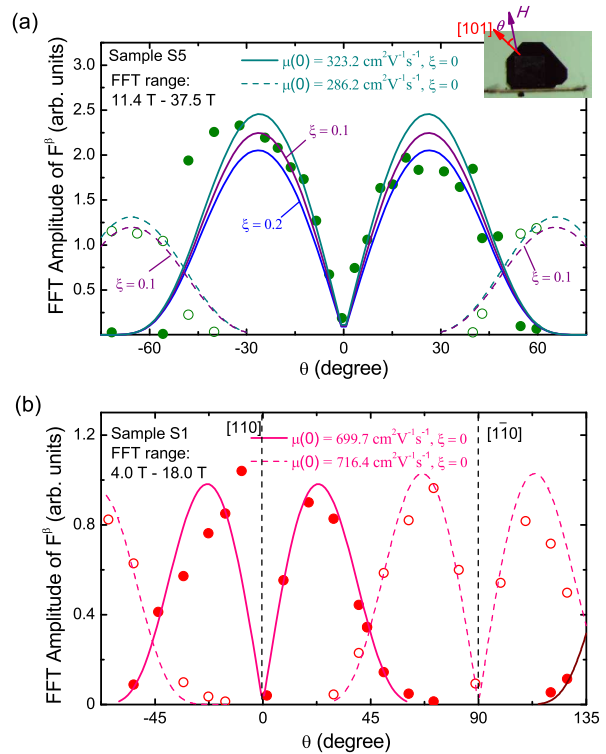}
\caption{Angular dependence of oscillation amplitudes. (a) The amplitude of peak $\beta$ in the FFT spectra of $M_{eff}$, plotted against the tilt angle $\theta$ between $H$ and [101] direction. Inset shows how $\theta$ is defined. The solid (fitted by the solid line) and hollow (fitted by the dashed line) symbols denote the amplitude of $F^\beta$ assumed to come from surfaces (101)($\bar{1}0\bar{1}$) and (10$\bar{1}$)($\bar{1}$01), respectively. The fittings (dark cyan) are made using the 2D LK model in Eq. \ref{fitting} and yield $\xi$ = 0. Results with $\xi$ = 0.1 (purple) and 0.2 (blue) are also shown for comparison. (b) Amplitude analysis using the same model applied on the data of an old SmB$_6$ samples, S1, measured up to 18 T. Data are extracted from Fig. S3 in Ref. \cite{li2014two}.
}
\label{FigAmpl}
\end{figure}

The angular dependence of the FFT amplitude of dHvA oscillation branch $F^\beta$ is plotted in Fig. \ref{FigAmpl}(a). Here we use a tilt angle $\theta$ with different definition: the angle between $H$ and the [101] direction of the crystal, as depicted in the inset of Fig. \ref{FigAmpl}(a). As expected for an effective magnetization extracted from magnetic torque data, this amplitude will drop to zero if the total magnetization vector is parallel/antiparallel to the applied magnetic field which result in no ``effective" component to be detected. In Fig. \ref{FigAmpl} this is shown to happen at both $H \parallel$ [101] and $H \perp$ [101], once again suggesting that pocket $\beta$ is related to the (101) planes. With the magnetic field rotated approximately 60$^\circ$ away from the symmetric axis [101], the FFT amplitude is reduced by a factor of 20-100, consistent with the behavior of the 2D topological surface state in Bi$_{1-x}$Sb$_x$ \cite{taskin2009quantum} (A detailed comparison is provided in Appendix \ref{chap06}). Beyond this tilt angle, the FFT can hardly pick up the oscillation signal from the corresponding Fermi pocket.

The most interesting feature of the $\theta$-dependent FFT amplitude is the absence of four-fold symmetry corresponding to the cubic crystal structure. That is, from [100] to [00$\bar{1}$] direction ($\theta < -45^\circ$) and from [001] to [$\bar{1}$00] direction ($\theta > +45^\circ$), the amplitude of $F^\beta$ is obviously smaller than that between [100] to [001] ($-45^\circ < \theta < +45^\circ$). The broken four-fold rotational symmetry in (010) plane strongly suggests that the oscillation frequency may not have a bulk origin. Also, this inequivalence between axes [101] and [10$\bar{1}$] shows sample dependence. In Fig. \ref{FigAmpl}(b) we summarize the angle-resolved dHvA amplitudes in another SmB$_6$ single crystal, S1, measured with the same experimental set-up. The two-fold feature in Fig. \ref{FigAmpl}(b) is much weaker, suggesting the symmetry breaking is more related to the sample instead of the cantilever magnetometry setup.

A reasonable interpretation of this symmetry breaking is the surface origin of $F^\beta$. In our magnetic torque measurement, the signal from two parallel surfaces (e.g. (101) and ($\bar{1}0\bar{1})$) will be picked up simultaneously, and in the field rotation in (001) plane, we can obtain the magnetic response from two individual sets of surfaces, which are perpendicular to each other. These two sets of surfaces are prone to have different plane impurity densities and subsequently differed carrier scattering rates that can apparently affect the amplitude of the quantum oscillation (Appendix \ref{chap06}). Here, we analyze the angular dependence of dHvA amplitude of $F^{\beta}$ using a 2D LK model \cite{shoenberg2009magnetic}:

\begin{equation}
\label{fitting}
\Delta M_\perp(\theta) \propto \frac{\sin \theta}{\cos^2 \theta} \exp(-\frac{\pi}{\mu(0) B \cos\theta}) \exp(-\xi \cos\theta)
\end{equation}

where $M_\perp$ is the effective magnetic moment picked up in torque measurement, $\mu(0)$ is the carrier mobility at $\theta = 0$: $\mu(0) = e\tau_{s}(0)/m^*(0)$, and $\xi = \pi\lambda B/\mu(0)$. This model takes the 1/$\cos\theta$ anisotropy of the cyclotron mass m$^{*}(\theta)$ as the main contribution to the angular dependence of oscillation amplitude. The comprehensive simplification process of this model, in which we consider the anisotropy of each term in the LK formula in 2D case, is presented in Appendix \ref{chap06}. We also applied Eq. \ref{fitting} to the dHvA data of topological surface state as well as bulk state in Bi$_{1-x}$Sb$_x$ reported in Ref. \cite{taskin2009quantum}, the fittings shown in Appendix Fig. \ref{FigBiSb}(a) and (b) provide strong evidence that Eq. \ref{fitting} is a valid model in describing the two dimensionality of electronic states and can effectively track the difference in the angle-dependent quantum oscillation amplitude between 2D and 3D system.

As shown by the fittings in Fig. \ref{FigAmpl} and Appendix Fig. \ref{FigBiSb}(c), the two-fold symmetry in FFT amplitude can be well described by a difference of carrier mobility $\mu(0)$ on the two perpendicular sets of surfaces. By assuming a reduction on $\mu(0)$ of $\simeq 11.5\%$ (Fig. \ref{FigAmpl}(a)), the nearly 50$\%$ amplitude suppression is reproduced even for a $25\%$ larger surface area (estimated from the sample geometry in the inset of Fig. \ref{FigTorque}(a)). This result substantially supports the 2D nature of oscillation branch $\beta$, as even an elongated ellipsoidal FS shows an evident deviation at high tilt angle (Appendix \ref{chap06}). Furthermore, the effective fitting by using zero or small value of $\xi$ reveals an ignorable Zeeman attenuation of the scattering rate (Appendix \ref{chap06}).

The mobility difference is much larger between different samples, as in sample S2 $\mu(0)$ (Appendix Fig. \ref{FigBiSb}(c)) is more than three times as large as in sample S5 (Fig. \ref{FigAmpl}(a)). This sample-dependent behavior is more likely due to the varied surface impurity level within samples. As we know, the scattering on the surface of SmB$_6$ is highly related to the surface disorder \cite{dzero2015nonuniversal}, and the dephasing length is differed by several hundred percent in different samples \cite{thomas2016weak}. Also, the carrier mobility $\mu$(0) we obtained from the fitting is only 50$\%$-70$\%$ of that calculated from the Dingle plot (see Appendix \ref{chap07}). This discrepancy can be addressed to the complicated electron scattering mechanism in SmB$_6$ which remains an enigma to be solved by further studies. We point out that the mobility attained by our dHvA amplitude fitting and Dingle analysis yield the same order of magnitude and are both much higher that those obtained from transport experiments \cite{syers2015tuning,wolgast2015magnetotransport}.

\begin{figure*}[htbp!]
\includegraphics[width=0.7\textwidth]{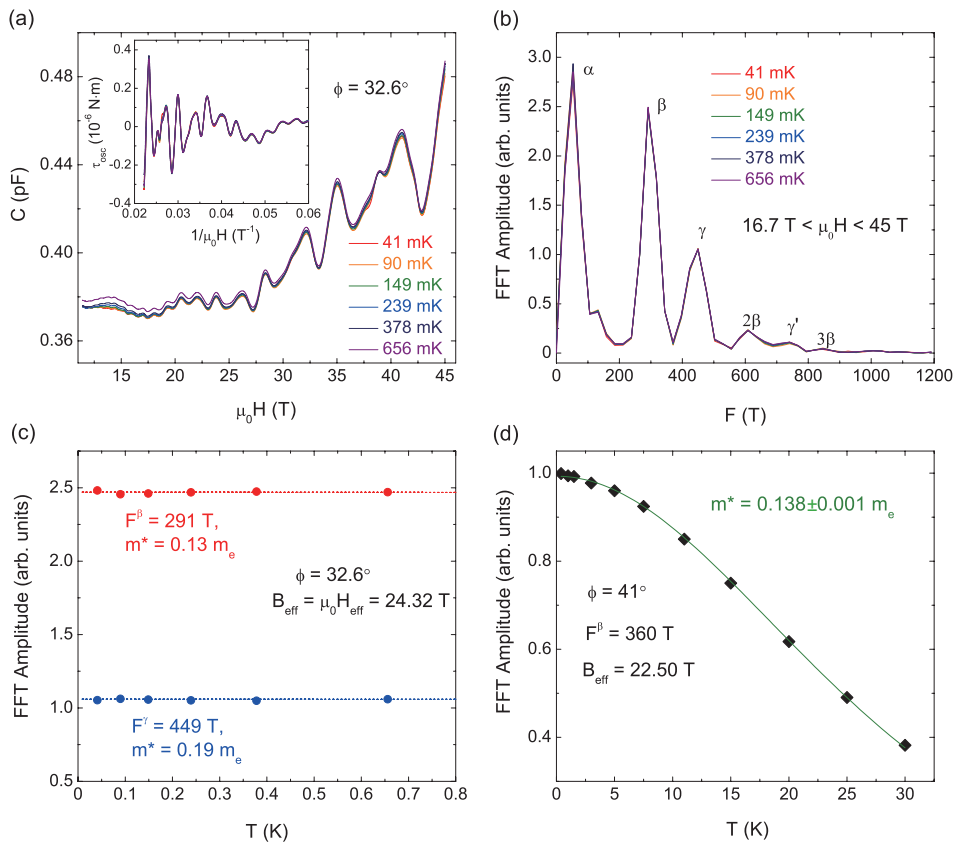}
\caption{Temperature dependence of oscillation amplitudes. (a) Capacitance signals in Sample S5 measured up to 45 T at different temperatures ranged from 41 mK to 656 mK. The tilt angle for this data set is $\phi$ = 32.6$^\circ$. Inset: The oscillatory part of magnetic torque extracted from the capacitance curves, as a function of inverse magnetic field. A polynomial background is subtracted. (b) The FFT amplitude curves of magnetic torque shown in the inset of (a) in a field range between 16.7 T and 45 T. (c) The FFT amplitudes of $F^{\beta}$ and $F^{\gamma}$ plotted as a function of temperature. Dashed lines are fittings based on Lifshitz-Kosevich formula with effective mass $m^{*}$ = 0.13 $m_{e}$ and 0.19 $m_{e}$ for oscillation branches $\beta$ and $\gamma$, respectively. (d) Temperature dependence of the FFT amplitude of $F^{\beta}$ tracked up to 30 K at $\phi = 41^\circ$. Fitting by LK formula yield an effective mass of 0.138 m$_{e}$.
}
\label{FigTemp}
\end{figure*}

Finally, the fitting based on the LK formula in Fig. \ref{FigAmpl} demonstrated again the validity of the Fermi Liquid theory in SmB$_6$. The mysterious sudden enhancement of quantum oscillation amplitude in SmB$_6$ reported by Tan et al. \cite{tan2015unconventional} has attracted much attention as a rare 3D example of the deviation from LK formula, which has been suggested as a reflection of unconventional quantum oscillation \cite{knolle2015quantum,zhang2016quantum,Pal,exciton,Pal2017}. In our dHvA studies down to 40 mK, however, such behavior is not repeated, even though we applied up to 45 T magnetic field, stronger than that used in Ref. \cite{tan2015unconventional}. In Fig. \ref{FigTemp} we summarized the temperature dependence of dHvA oscillations at $\phi = 32.6^\circ$: the capacitance $C(B)$ (Fig. \ref{FigTemp}(a)), the oscillatory magnetic torque $\tau_{osc}$ (inset of Fig. \ref{FigTemp}(a)) and the FFT curves of $\tau_{osc}$ (Fig. \ref{FigTemp}(b)) all show almost no discernible difference at varied temperatures between 41 mK and 656 mK. None of the dHvA branches we resolved have any abnormal enhancement in this temperature range. Fig. \ref{FigTemp}(c) summarizes the evolution of the FFT amplitude. The light effective masses for both $F^\beta$ and $F^\gamma$ together with the low $T$ give a temperature damping factor $R_T$ very close to 1. As a result, the oscillation amplitude is generally a constant with the relative change almost ignorable. We also tracked the evolution of the FFT amplitude of $F^\beta$ from 350 mK up to 30 K, at a tilt angle of $\phi = 41^\circ$, as shown in Fig. \ref{FigTemp}((d). Assuming that the 2D LK formula the damping factor related to temperature is the same as that for the 3D case \cite{singleton2000studies}, the overall behavior can be fitted by LK formula with an effective mass of 0.138 $m_{e}$, consistent with our former report \cite{li2014two}. The weak temperature dependence of dHvA amplitude below 300 mK is confirmed by measurements in different samples and at various tilt angles, which is shown in Appendix Fig. \ref{FigAddTemp}.

The distinct physical phenomena observed in flux-grown and FZ-grown SmB$_6$ are quite intriguing and confusing. The FZ growth are reported to induce a small portion of Sm vacancies in SmB$_6$ single crystal \cite{phelan2016chemistry,phelan2014thermo,Valentine}. Subsequently, a slight difference on the Sm valence on the surface can be established, which in turn modifies the Kondo interaction near the surface \cite{alexandrov2015kondo,erten2016kondo}. Evidences of incomplete Kondo coupling in SmB$_6$, especially at the vicinity of the surface, have been discovered \cite{wolgast2015magnetotransport,nakajima2016one,Valentine,Biswas}.  Actually, in a topological Kondo insulator, both light and heavy surface states can be supposed to appear by varying the detailed band parameters \cite{feng2016dirac}. Besides this valence variation scenario induced by non-stoichiometry in the framework of Kondo physics, another interpretation of the inconsistencies between our results and those reported in FZ-grown crystals by Tan \textit{et al.} \cite{tan2015unconventional} is attributed to different disorder levels. Recent theory points out a topological Kondo insulator can be restored from an exotic Skyrme insulator, in which dHvA oscillation is contributed by scalar charge neutral particles, by introducing a certain degree of disorder \cite{erten2017skyrme}. The absence of high-frequency oscillations in our sample can also be attributed to the higher scattering rate induced by the impurity/defects. Furthermore the deviation of LK behavior in FZ-grown samples was taken as evidence for exotic nature of oscillations \cite{tan2015unconventional}, we note that a similar double-step feature was observed in topological nodal semimetal ZrSiS ~\cite{ZrSiS}, in which the Fermi surface nesting leads to two LK behavior with two different effective masses. This is a more realistic origin than the various unconventional quantum oscillation models \cite{knolle2015quantum,zhang2016quantum,Pal,Pal2017,exciton,baskaran2015majorana,erten2017skyrme}. Overall, the topological surface state explanation is still the most natural one for all of our observations of the magnetic quantum oscillations in flux-grown SmB$_6$.

In summary, magnetic torque data of Kondo insulator SmB$_6$ measured in intense magnetic field at dilution refrigerator temperature has been investigated comprehensively. The amplitude of main dHvA oscillation branch $\beta$, which shows 1/$\cos\phi$ angular dependence in frequency, displays a broken four-fold symmetry with field rotating in crystalline (010) plane. The angle-dependent oscillation amplitude can be fitted by a standard 2D LK model with respected to each sets of (101) planes. The carrier scattering rate obtained from the fittings is differed significantly between samples as well as surfaces on the same sample. The dHvA oscillations are also fully saturated at low temperature, implying small carrier masses. Our results indicate that multiple 2D light electron states are existed on the surfaces of SmB$_6$.

\addcontentsline{toc}{section}{ACKNOWLEDGMENTS}
\section*{ACKNOWLEDGMENTS}

This work is mainly supported by the National Science Foundation under
Award No. DMR-1707620 (magnetization measurement), by the Office of
Naval Research through the Young Investigator Prize under Award No.
N00014-15-1-2382 (electrical transport characterization), and the
National Science Foundation Major Research Instrumentation award under
No. DMR-1428226 (supports the equipment of the thermodynamic and
electrical transport characterizations). C. S. and X.H.C. thank the
National Key R \& D Program of the MOST of China (Grant No.
2016YFA0300201). The development of the torque magnetometry technique in intense magnetic
fields was supported by the Department of Energy under Award No.
DE-SC0008110. Some experiments were performed at the National High
Magnetic Field Laboratory, which is supported by NSF Cooperative
Agreement No. DMR-084173, by the State of Florida, and by the DOE. The
experiment in NHMFL is funded in part by a QuantEmX grant from ICAM
and the Gordon and Betty Moore Foundation through Grant GBMF5305 to
Dr. Ziji Xiang, Tomoya Asaba, Colin Tinsman, Lu Chen, and Dr. Lu Li.
We are grateful for the assistance of Tim Murphy, Glover Jones,
Hongwoo Baek, and Ju-Hyun Park of NHMFL. T.A. thanks the Nakajima
Foundation for support. B.J.L. acknowledges support by the National
Science Foundation Graduate Research Fellowship under Grant No.
F031543.

\appendix

\section{Materials and methods}
\label{chap01}

SmB$_6$ single crystals were grown by the aluminum (Al) flux method \cite{canfield1992growth}. The chunks of Sm (99.95$\%$), the powder of Boron (99.99$\%$) and Al (99.99$\%$) were mixed together with a mass ratio of 1:6:400, and then loaded into an alumina crucible. The entire mixture was heated to 1550 $^\circ$C and then stayed at this temperature for 2 days before cooled down to 600 $^\circ$C at 5 $^\circ$C/h. During all the preparing and heating progress, the mixture was kept in the Argon gas. After cooled to room temperature, the samples with Al flux were soaked in the dense NaOH solution to remove the Al flux, and then washed by dilute HNO$_3$ solution. The samples were characterized by X-ray diffraction(XRD) to determine the orientation.
Upon cooling from room temperature to $^3$He temperature, the resistivity of our SmB$_6$ samples is enhanced by 4-5 orders of magnitude, and a resistive plateau shows up below 3.5 K \cite{chen2015magnetoresistance}. Data discussed in this work were mainly taken from SmB$_6$ sample S5, which has a size of 2.1$\times$1.6$\times$1.2 mm$^3$ and hosts large (100) and (101) surfaces. A smaller sample labeled as S6 was also measured but the results shown no significant differences and the signal quality is lower due to the smaller quantum oscillation amplitudes. All the samples we used are as-grown single crystals without cleaving or polishing.

The high magnetic field torque magnetometry measurements were carried out using the capacitance method in National High Magnetic Field Laboratory (NHMFL), Tallahassee. The SmB$_6$ samples were glued to a beryllium-copper cantilever with the thickness of 0.025 mm. Variation in the capacitance between the cantilever and a fixed gold film reflects the bending of cantilever, from which the magnetic torque can be obtained. The set-up with sample S5 attached is shown in Fig. \ref{FigTorque}(a). Such devices were put into a rotator and then loaded into a dilution fridge in a hybrid magnet which can apply magnetic field up to 45 T. The cantilevers were rotated with magnetic field in the crystalline (010) plane.

We measured the capacitance change with sweeping magnetic field via two methods. The frequently used Andeen-Hagerling AH2700A digital capacitance bridge usually has a noise level of 10$^{-4}$ pF in our experimental environment, and the automatic balancing is slow, which means it may not be suitable to pick up the weak high-frequency dHvA oscillations in SmB$_6$. As an alternative, we chose the General Radio analog capacitance bridge combined with the Stanford Research SR124 analog lock-in amplifier. By balancing the starting capacitance $C_0$ manually and reading the voltage change during field sweeping, we can achieve a better resolution with the noise level reduced by one order of magnitude. Also this allows for a continuous reading of the cantilever response.

\section{Exclusion of the flux-induced quantum oscillation}
\label{chap02}

Extrinsic quantum oscillations introduced by Al flux trapped inside the sample has been reported in CaB$_6$ \cite{terashima2000ferromagnetism}. The existence of epitaxially oriented Al flakes in the flux-grown SmB$_6$ single crystals has also been confirmed, with a percentage of 2-4 wt$\%$ \cite{phelan2016chemistry}. However, the dHvA amplitude of single-crystalline aluminum is known to be $\Delta \tau \simeq 3.5 \times10^{-7} N \cdot m$ at 4.2 K under $B$ = 2 T, for the strongest oscillation branch $\gamma_5$ in a sample with the mass of 45 mg \cite{larson1967low}. Considering of the reported effective mass $m^*/m_e$ = 0.18 and Dingle temperature $T_D$ =0.8 K for this band in aluminum, there is an amplifying factor of approximately 60x in torque for the condition of $B$ = 40 T and $T$ = 45 mK. It means the weight of incorporated Al, if it contributes to all the dHvA signals shown in Fig. \ref{FigTorque}(b), should be $\sim$ 2 mg. Since the weight of SmB$_6$ sample S5 is 12.9 mg, the amount of co-crystallized Al could be as large as 15 wt$\%$ (58.4 mol$\%$) which is much larger than that revealed by X-ray diffraction study \cite{phelan2016chemistry}. Consequently, the dHvA patterns in Fig. \ref{FigTorque}(b) is more likely to be intrinsic.

\begin{figure}[htbp!]
\centering
\includegraphics[width=0.48\textwidth]{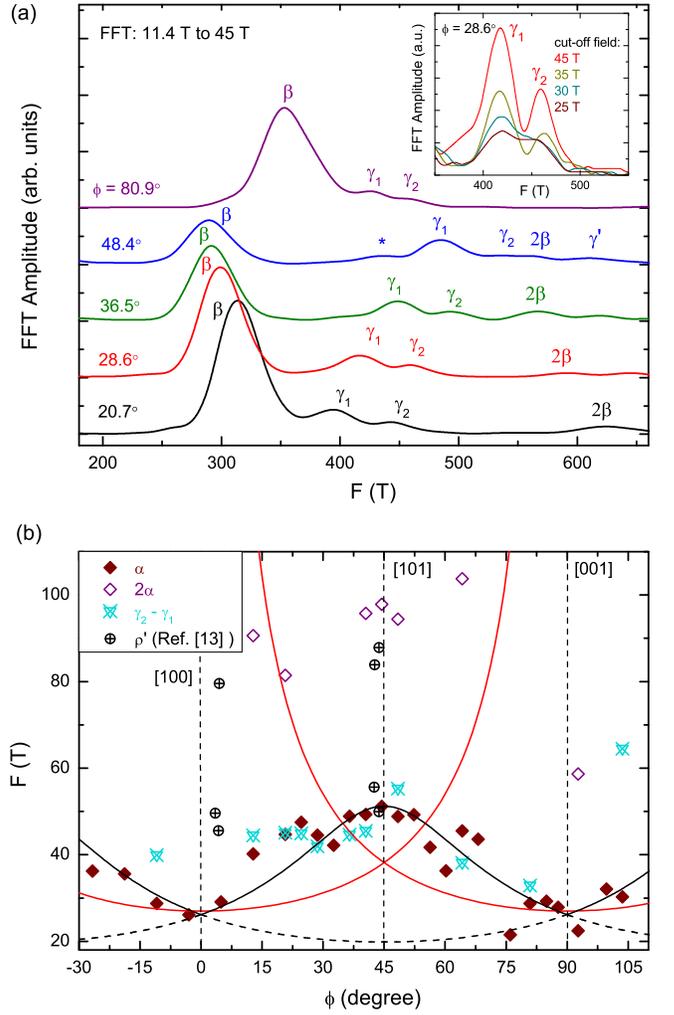}
\caption{(a) The FFT spectra of $M_{eff}$ for various tilt angles. Two peaks on the high-frequency side of the major peak $\beta$ are labeled as $\gamma_1$ and $\gamma_2$, respectively. One unknown small feature on the $\phi$ = 48.4$^\circ$ at 435 T is marked by a star. Inset: The FFT spectrum at $\phi$ = 28.6$^\circ$, between 11.4 T and a varied cut-off field. The split of peak $\gamma$ is more evident as the cut-off field becomes larger. (b)The angle dependence of lowest frequency oscillation branch $\alpha$ and its second harmonic, plotted together with the frequency interval between split $\gamma_1$ and $\gamma_2$ and data of frequency $\rho'$ extracted from Ref. \cite{tan2015unconventional}. The red curves are fittings using a 2D Fermi surface (FS) model, whereas the black curves are 3D fittings based on an ellipsoid FS. For the 3D model, we follow the previous report in which the geometry of the FSs contributing to $\rho'$ are figured out to be small ellipsoidal pockets with their long axes along the cubic [101] directions \cite{tan2015unconventional}. The fitting (black solid line) gives a ratio of long axis (along [10$\overline{1}$]) : short axis (along [101]) = 2.58. However, the other set of equivalent FSs with long axis along [101] (black dashed line) are absent from our data.
}
\label{FigLowFreq}
\end{figure}

There are extra evidences against the flux-induced extrinsic quantum oscillations. First, there is no two-peak feature in any of the $\gamma$ branches in the dHvA oscillations of aluminum \cite{ashcroft1963fermi,larson1967low}, where as such a peak split has been observed in our flux-grown SmB$_6$ samples (Fig. \ref{FigLowFreq}(a)). Second, there are evident discrepancies between the angular dependence of oscillation frequencies in SmB$_6$ and Al as reported before \cite{li2014two}. Also, the resemblance of dHvA frequencies in flux grown samples and floating-zone grown samples (which are grown free of Al, see Appendix \ref{chap05}) supports that the oscillations are intrinsic.

\begin{figure*}[htbp!]
\centering
\includegraphics[width=0.8\textwidth]{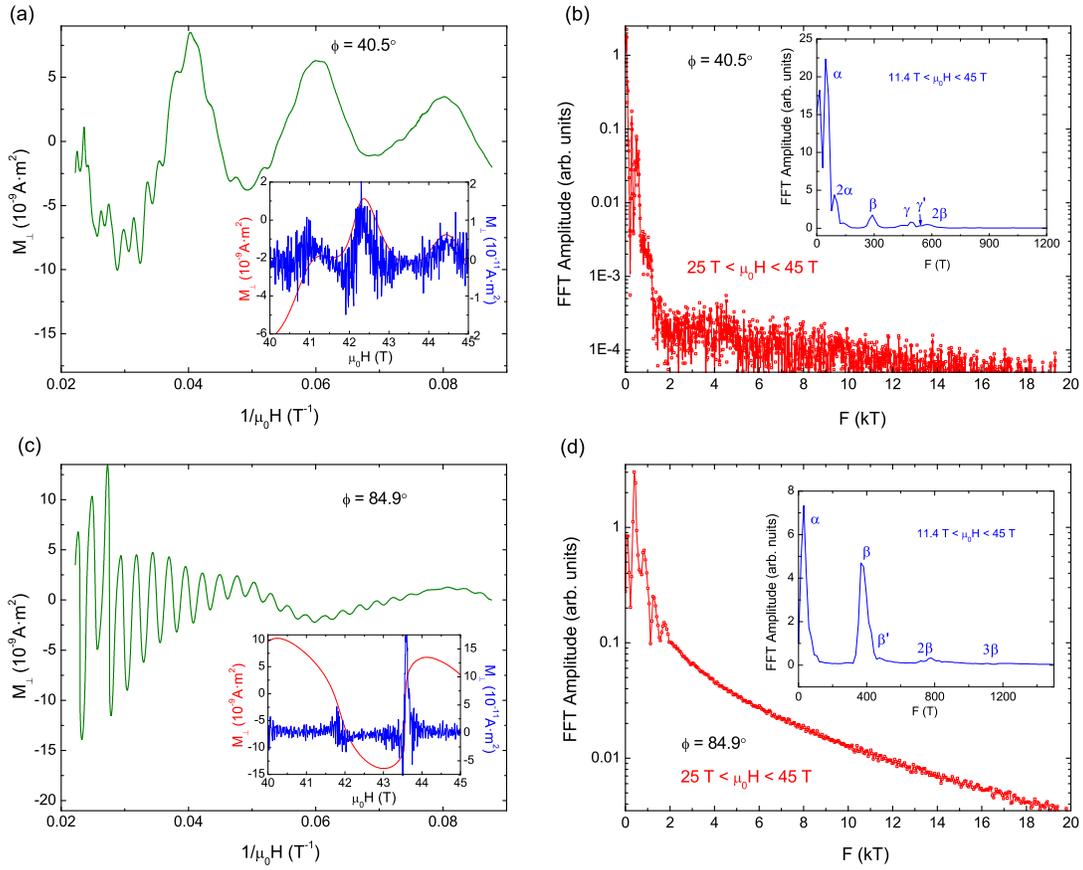}
\caption{Absence of high-frequency oscillations. The oscillatory effective magnetization $M_{eff}$ as a function of inverse magnetic field at (a) $\phi$ = 40.5$^\circ$ (c) $\phi$ = 84.9$^\circ$. Polynomial backgrounds are subtracted from the raw data. Insets: the low-frequency component of $M_{eff}$ obtained by a low-pass FFT filter with threshold frequency 2 kT (red) and the residual component after subtracting the low-frequency oscillation from $M_{eff}$ (blue), both shown in a magnetic field range between 40 T and 45 T. The FFT results of the oscillatory part of $M_{eff}$ at (b) $\phi$ = 40.5$^\circ$ (d) $\phi$ = 84.9$^\circ$ between 25 T and 45 T. No oscillation frequency higher than 2 kT can be resolved from the background at both tilt angles. Insets show the low-frequency peaks with $F <$ 1200 T in a full range FFT from 11.4 T to 45 T.
}
\label{FigHighFreq}
\end{figure*}

\begin{figure}[htbp!]
\centering
\includegraphics[width=0.46\textwidth]{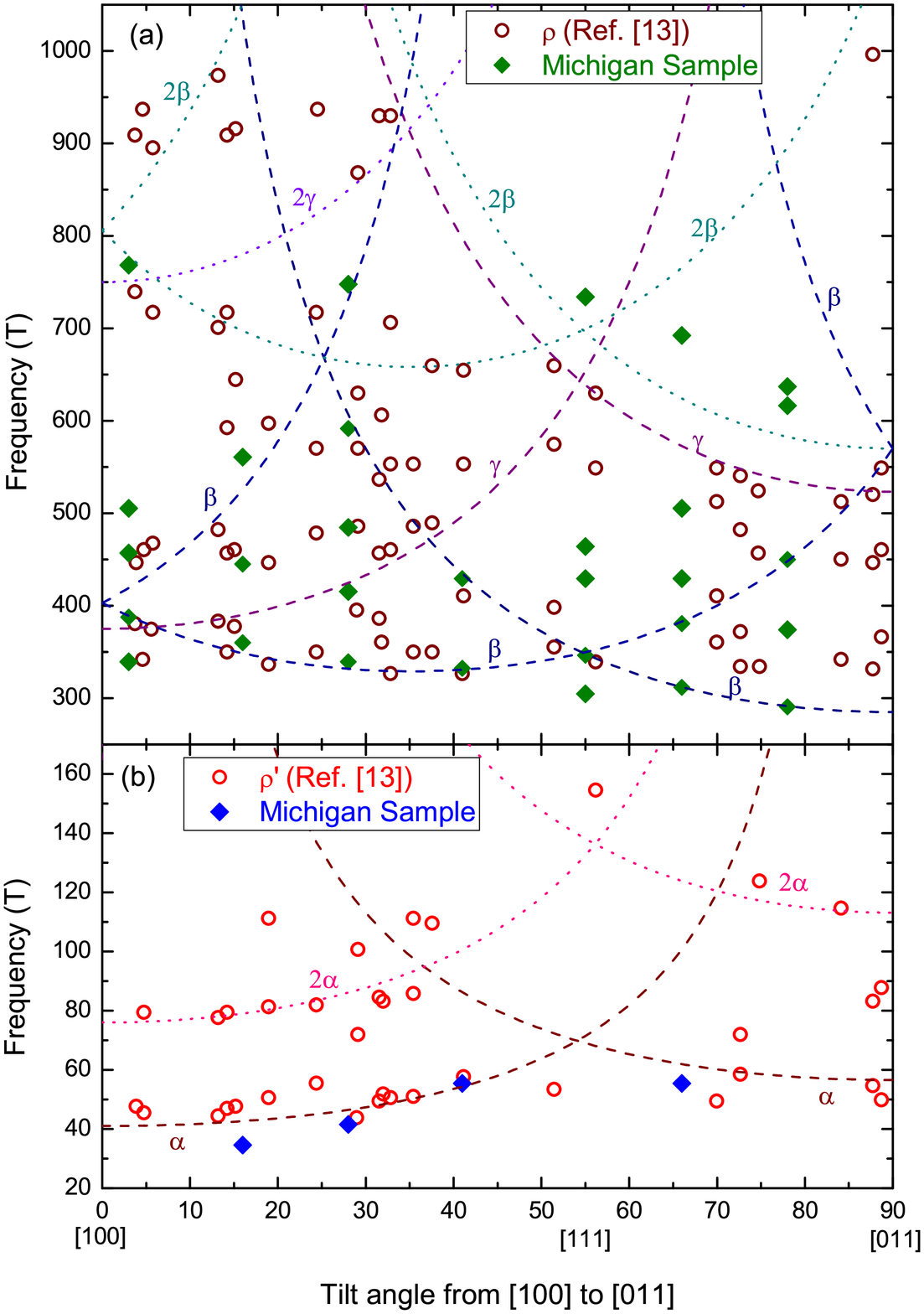}
\caption{A comparison of the FFT peaks resolved from the magnetic torque data in the floating-zone grown SmB$_6$ single crystals, extracted from Ref. \cite{tan2015unconventional}, and the Al-flux-grown samples studied in Ref. \cite{li2014two}. In panel (a) and (b) the FFT peaks $\rho$ and $\rho'$ in Ref. \cite{tan2015unconventional} are plotted, respectively, together with the dHvA oscillation features in the flux-grown samples that within the same frequency range. In both data sets, the magnetic field is rotated from crystal [100] axis towards [011] axis. Dashed lines are fittings by the 2D cylinder FS model \cite{li2014two}, and dot lines are fittings of the harmonics.
}
\label{FigCompare}
\end{figure}

\section{The splitting on $\gamma$ and the low frequency oscillation $\alpha$}
\label{chap03}

There are two adjacent FFT peaks at the location of branch $\gamma$, as shown in Fig. \ref{FigLowFreq}(a), with a frequency interval of 43$\pm$5 T for most of the angles. We label the peak with lower frequency $\gamma_1$ and the high-frequency one $\gamma_2$. Such phenomenon is also presented in our old data, though under lower magnetic field (up to 18 T) the splitting is less clear \cite{li2014two}. The inset of Fig. \ref{FigLowFreq}(a) gives a comparison of the FFTs with different end points, and the splitting appears to be unambiguous only for a cut-off field higher than 30 T. The origin of this observation is unclear. If the FS $\gamma$ is a quasi-two-dimensional one and has a small periodic warping along $k_z$ direction, two extrema of the FS cross-sectional area can result in two quantum oscillation frequencies that are close to each other \cite{yamaji1989angle,coldea2008fermi}. In this scenario, however, the two extreme areas have different curvature in the angle dependence and can cross together at certain angles. These expectations are absent in our data (see the triangle symbols in Fig. \ref{FigTorque}(c)). We suggest a more plausible explanation that the two peaks $\gamma_1$ and $\gamma_2$ respectively come from two sets of electron orbits on (100) surfaces. Since the pockets $\alpha$ and $\gamma$ have identical symmetric axes along [100], one possibility is that the split is originated in a magnetic breakdown between them, i.e., $\gamma_2$ = $\gamma_1$+$\alpha$.

We notice that the interval between the two split $\gamma$ branches show insignificant change with varying field directions (Fig. \ref{FigTorque}(c)). As exhibited in Fig. \ref{FigLowFreq}(b), the small angle dependence of $\gamma_2$-$\gamma_1$ almost coincides with that of $F^{\alpha}$, which raises the possibility of magnetic breakdown between orbits $\gamma_1$ and $\alpha$. Taking into account that the splitting of $\gamma_2$ is only apparent under high magnetic field (inset of Fig. \ref{FigLowFreq}(a)), magnetic breakdown is a rather promising explanation. With the orbit area differed by one order of magnitude, however, magnetic breakdown is unlikely to happen if $\gamma$ and $\alpha$ are centered at the same momentum position in the Brillouin zone. It is probable that one of the two orbits is located away from the high-symmetry points in momentum space, but in this case additional frequencies like $\gamma_1$+$\textit{n}$$\alpha$ or $\alpha$+$\textit{n}$$\gamma_1$ are supposed to be observed according to the crystal symmetry. Actually, they are absent in our FFT spectra. At this stage, the reason of the splitting and the locations of orbits $\gamma$ and $\alpha$ remain unclear.

The smallest orbit $\alpha$ resolved from the FFT of the torque curves has an indeterminate geometry. In Fig. \ref{FigLowFreq}(b) we try to simulate the angle dependence of $F^{\alpha}$ by both 2D cylinder model and 3D elongated ellipsoid model. The 2D FS model can not exactly follow the frequency increase from [100] direction to [101] direction, whereas 3D fittings based on an ellipsoidal FS with long axis along crystalline [10$\overline{1}$] direction can track the data fairly well. Considering the cubic symmetry of the crystal structure, equivalent FSs with the long axis along [101] should exist as well. However, they are totally absent in our FFT analysis. Besides, the FFT peaks with higher frequencies (hollow symbols in Fig. \ref{FigLowFreq}), which are identified as the 2nd harmonics here, only show up in the vicinity of [101] direction where the oscillation amplitude of $\alpha$ is the strongest. This behavior suggests those peaks in the range of 80-110 T are harmonics in nature instead of the diverging $F^{\alpha}$ from one of the same set of FSs with the symmetry axis 90$^\circ$ away (as $\beta'$ and $\gamma'$ plotted in Fig. \ref{FigTorque}(c) as well as in Ref. \cite{li2014two}). We also added the data reported in Ref. \cite{tan2015unconventional} to Fig. \ref{FigLowFreq}(b), though considering the different sample rotation direction in the two studies we only select the angles close to 0$^\circ$ ([100]) and 45$^\circ$ ([101]). According to Tan \textit{et. al.}, the smallest orbit $\rho'$ is assigned to a small ellipsoid inside the ``neck" connecting the large FSs \cite{tan2015unconventional}. However, judging from the consistency in Fig. \ref{FigLowFreq}(b) it is arguable that $\rho'$ branch is corresponding to our $\alpha$ and its second harmonic.

\section{Absence of high-frequency dHvA oscillations}
\label{chap04}

Oscillation branches with frequencies higher than 2 kT, while have been detected in the floating-zone furnace grown SmB$_6$ single crystals \cite{tan2015unconventional}, are completely missing in our measurement. Actually the frequencies between 1 kT and 2 kT are already rather weak in our FFT spectra. In Fig. \ref{FigHighFreq} we show the analysis for two field orientations, i.e., $\phi = 40.5^\circ$ at which the low-frequency branch $\alpha$ is strong but $\beta$ and $\gamma$ are relatively weak, and $\phi = 84.9^\circ$ where branch $\beta$ has a large spectral weight. At both angles the capacitance was measured by analog capacitance bridge. The oscillatory part of $M_{\perp}$, the perpendicular component of magnetization, is dominated by the ``slow" dHvA oscillations (Fig. \ref{FigHighFreq}(a) and (c)). After subtracting those ``slow" components (with $F <$ 1200 T), the residual magnetization term is barely noise with the amplitude approximately 10$^{-11}$ A$\cdot$m$^2$, that is, 0.1-0.2$\%$ of the total oscillatory $M_{\perp}$ (see the inset of Fig. \ref{FigHighFreq}(a) and (c)). No periodic small wiggles can be isolated that can indicate the existence of fast quantum oscillations. On the FFT of $M_{\perp}$, everything with frequency higher than 2 kT sinks into the background noise that is roughly 1/1000 of the main peak height and no features can be resolved, even when we take the FFT in a high field range (Fig. \ref{FigHighFreq}(b) and (d)).

\section{Comparison between the dHvA frequencies in flux-grown and floating-zone-grown samples}
\label{chap05}

In Fig. \ref{FigCompare} we provide a comparison of the results reported by two groups on the dHvA frequencies resolved from SmB$_6$ single crystal samples grown via different approaches, i.e., Al-flux crystals studied by Li \textit{et. al.} \cite{li2014two} and in floating-zone crystals studied by Tan \textit{et. al.} \cite{tan2015unconventional}. Measurements are taken with the same technique and experimental conditions \cite{li2014two,tan2015unconventional}. As is confirmed by Fig. 5, the high-frequency components with $F >$ 2 kT are not detected in the flux-grown samples. On the other hand, the low-frequency FFT peaks in these two data sets are generally comparable, with data points falling into the same frequency range and the overall trend of angular dependence also show some similarities. It highly suggests that (i) those dHvA frequencies are intrinsic in SmB$_6$ (ii) the two works are looking into the quantum oscillations from the same electronic states.

\section{Angle dependence of dHvA oscillation amplitude in 2D electron system}
\label{chap06}

For a 2D electron system, given the condition of $F/B \gg 1$, the amplitude of longitudinal magnetization quantum oscillations can be approximately described by a 2D LK expression \cite{harrison1996numerical,harrison1995haas,singleton2000studies,ramshaw}:
\begin{equation}
M_\parallel = -A\sum_{p=1}^{\infty}(\frac{1}{2\pi p})R_TR_DR_S sin[2\pi p(\frac{F}{B}-\gamma)]
\label{2DLK}
\end{equation}
where $R_T = X_{p}/\sinh(X_{p})$, $X_{p} = 2\pi^2 p k_B m^* T/e \hbar B$, $R_D = \exp(-2\pi^2 k_B m^* T_{D}/e \hbar B)$, $T_D = \hbar/2\pi k_B \tau_{Q}$ the Dingle temperature, $A$ a parameter proportional to quantum oscillation frequency $F$, and $R_S = \cos(p\pi g m^*/2m_e)$ the spin-splitting factor. Here $m_e$ is the free electron mass and $\tau_{Q}$ the quantum oscillation relaxation time. In this work we study the magnetic torque, and the effective magnetization extracted from the torque signal is $M_{eff}$ = $M_{\perp}$. According to Ref. \cite{shoenberg2009magnetic},
\begin{equation}
M_\perp = -\frac{1}{F} \frac{\partial F}{\partial \theta} M_\parallel
\label{Mperp}
\end{equation}
which serves as the starting point of our dHvA oscillation amplitude analysis.

There are three independent parameters in Eq. \ref{Mperp} that are functions of the magnetic field tilt angle $\theta$: $F$ = $F(\theta)$, $m^*$ = $m^*(\theta)$ and $\tau_{Q}$ = $\tau_{Q}(\theta)$. The dHvA frequency $F$ appears in the universal coefficient of all the harmonics, $(1/F)(\partial F/\partial \theta) A$, in which $A$ $\propto$ $F$. According to our fitting in Fig. 2(b), $F^\beta(\theta)$ shows the typical characterization of a 2D FS, i.e., $F^\beta(\theta)$ $\propto$ 1/$\cos \theta$. Therefore we have:
\begin{equation}
\frac{1}{F} \frac{\partial F}{\partial \theta} A \propto \frac{\partial F}{\partial \theta} \propto \frac{\sin \theta}{\cos^2 \theta}
\end{equation}

The effective mass $m^*$ is included in all three amplitude factors of $R_T$, $R_D$ and $R_S$. For a parabolic-band system, the definition of effective mass is $m^*$ = $(\partial^{2} E/\partial k^{2})^{-1}$, whereas, for a linear dispersive 2D electron system such as the surface state of topological insulator, we can use the expression as follows \cite{novoselov2005two,ando2013topological}:
\begin{equation}
m^* = \frac{\hbar^2}{2 \pi}(\frac{\partial S(E)}{\partial E})_{E = E_{F}}
\label{Mass}
\end{equation}
where $S(E)$ is the cross sectional area of the 2D FS perpendicular to the field vector. Note that $S(E)$ $\propto$ $F$ varies as 1/$\cos \theta$ but the energy dispersion and carrier density will not change with the sample rotation in magnetic field, assuming the band dispersion relation is invariant along all directions and there is no magnetic-field induced modification in any of the bands. It means for 2D electron systems we have $m^*$ $\propto$ 1/$\cos \theta$ which is equally effective for the conventional parabolic and the Dirac-like band dispersion \cite{taskin2009quantum,taskin2010oscillatory}. We checked the anisotropy of effective mass in two SmB$_6$ single crystals (the FFT plots of these two samples can be found in Fig S3-S5 in Ref. \cite{li2014two}), and the results for pocket $\beta$ are $m^*$/$m_e$ = 0.124 (0.122) at $\theta$ = 14.6$^\circ$ and $m^*$/$m_e$ = 0.140 (0.147) at $\theta$ = 34.8$^\circ$ for sample S1(2). The relative offset of $m^*(\theta)$ regards to the expected 1/$\cos \theta$ behavior is therefore 1.9$\%$ and 4.2$\%$ for sample S1 and S2, respectively. Considering the sample misalignment and the LK fitting error, this result is quite reasonable, and we can also give an estimation of the effective mass at $\theta$ = 0$^\circ$: $m_0$ = $m^*(0) \simeq 0.120 m_e$.

To simplify the model, we take the Dingle factor $R_D$ as the only amplitude factor that is effectively influenced by the angle-dependent $m^*(\theta)$. This approximation is actually sensible owing to the following reasons. Giving the small value of $m_0$ for pocket $\beta$ and the low environment temperature, the angular dependence of $R_T$ is almost negligible. For $m_0$/$m_e$ = 0.12 and $T$ = 40 mK, the value of $R_T$ is very close to 1 with an offset smaller than 0.1$\%$ except for the angle range $|\theta-90^\circ| < 1^\circ$. In our measurement, the oscillation signal from $\beta$ branch on one surface cannot be detected with the field $\theta > 75^\circ$ away from the normal direction of the relating surface, as shown in Fig. \ref{FigAmpl}. Hence the factor $R_T$ can be safely treated as an constant in our fitting. We also left out the spin-splitting factor $R_S$ since we do not have a reliable estimation of the Land\'{e} factor $g$ for the light Fermi pocket $\beta$. If this pocket is a topological surface state, this factor will hardly play any role in affecting the quantum oscillation amplitude, because $R_S$ comes from the superposition of oscillations from the split Landau levels (i.e., spin-up and spin-down) \cite{shoenberg2009magnetic}. In a topological surface state there is no spin degeneracy at $k \neq 0$, accordingly the Zeeman effect shift the position of the Landau levels instead of causing the splitting \cite{taskin2009quantum,ando2013topological,fu2016observation}, consequently results in no reduction on the amplitude of oscillation.

\begin{figure}[htbp!]
\centering
\includegraphics[width=0.48\textwidth]{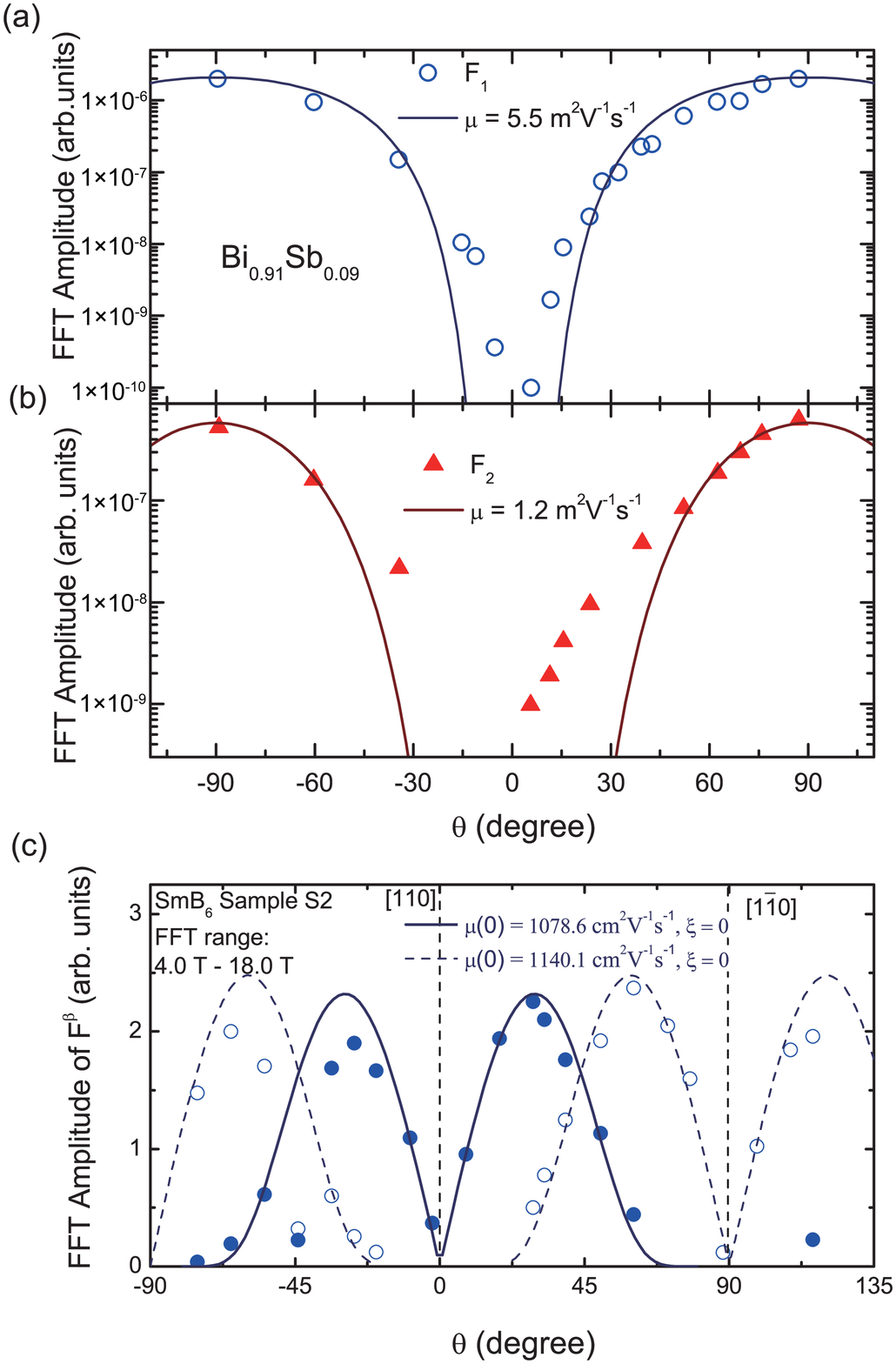}
\caption{The dHvA oscillation amplitudes of (a) 2D surface state on the bisectrix plane (b) 3D ellipsoidal bulk FS in Bi$_{1-x}$Sb$_x$, fitted by Eq. \ref{longitudinal}. Data points are extracted from Ref. \cite{taskin2009quantum}. Tilt angle $\theta$ is the angle between the magnetic field, which is rotated in the binary plane, and the crystal axis $C_3$. The carrier mobility in each panel is calculated from the parameters obtained in the same work. (c) Fitting of the angle-dependent amplitude of FFT peak $\beta$ in SmB$_6$ sample S2 by Eq. \ref{amplitude}, with parameter $\xi$ = 0. Definitions of $\mu_0$ and $\xi$ are the same as in Fig. \ref{FigAmpl}. Data are extracted from Fig. S4 in Ref. \cite{li2014two}.
}
\label{FigBiSb}
\end{figure}

The angular dependence of relaxation time $\tau_{Q}(\theta)$ is more complicated. We separate the magnetic field into two components, the in-plane field $H_{\parallel}$ = $H\sin\theta$ and the out-of-plane field $H_{\perp}$ = $H\cos\theta$. The first term is known to have no significant transport response from topological surface state in the absence of the hybridization between top of bottom surfaces \cite{lin2013parallel}, which can be completely neglected in our bulk single crystals. Theoretically, the in-plane magnetic field will only shift the position of surface Dirac point in momentum space \cite{garate2010magnetoelectric,zyuzin2011parallel} and cause a net in-plane spin polarization \cite{sulaev2015electrically}. A deformation of the Fermi pocket corresponding to the spin density redistribution is also suggested \cite{wang2015zeeman}. In all, the spin momentum-locking and the prohibition of backscattering in topological surface states is not destroyed in an in-plane magnetic field. The case is totally different for the second term, the out-of-plane component, which can break the time-reversal symmetry and lift the protection of the topological nontriviality. In this case, the back-scattering is reintroduced, and the electron-impurity scattering is enhanced by the Zeeman-energy-related spin-canting \cite{wu2015high}. The transport scattering rate takes the form as:
\begin{equation}
\frac{1}{\tau_{tr}} = \frac{1}{\tau_0}(1+\lambda B^2 \cos^2\theta)
\label{tau}
\end{equation}
$\lambda$ is a system parameter related to $g$-factor and Fermi energy.

Taking into account all the angle-dependent parameters discussed above, we can give a fitting model of the quantum oscillation amplitude of $M_{\perp}$ for the fundamental harmonic:
\begin{equation}
\label{amplitude}
\begin{split}
\Delta M_\perp(\theta) &= \frac{1}{F(\theta)} \frac{\partial F(\theta)}{\partial \theta} (\frac{A(\theta)}{2\pi})R_D(\theta) \\
 &\propto \frac{\sin \theta}{\cos^2 \theta} \exp(-2\pi^2 k_B m^* T_{D}/e \hbar B) \\
 &= \frac{\sin \theta}{\cos^2 \theta} \exp(-\frac{\pi}{\mu(0) B \cos\theta}) \exp(-\xi \cos\theta)
\end{split}
\end{equation}
where $\mu(0)$ is the carrier mobility at $\theta = 0$: $\mu(0) = e\tau_{s}(0)/m^*(0)$, and $\xi = \pi\lambda B/\mu(0)$. This expression is the same as Eq. \ref{fitting}. One needs to mention that the transport relaxation time $\tau_{tr}$ in Eq. \ref{tau} is different from the quantum oscillation relaxation time $\tau_{Q}$ we used in the Dingle factor, as the former one is more sensitive to backscattering \cite{liang2015ultrahigh}. Also, the field dependence of $\tau_{tr}$ in Eq. \ref{tau} is basically a weak-field approximation. As a simplification, in the fitting model Eq. \ref{amplitude} we assume that $\tau_{tr}$ and $\tau_{Q}$ share the same field dependence in the magnetic field range in our measurement. The detailed field effect on the scattering in topological surface states still needs further investigations.

To examine the validity of our model, we applied it to the quantum oscillation amplitudes in a well known topological insulator Bi$_{1-x}$Sb$_x$, reported by Taskin and Ando, in which both 2D and 3D FSs can be resolved in dHvA measurement \cite{taskin2009quantum}. Since the data in Ref. \cite{taskin2009quantum} was taken by SQUID magnetometer, the oscillations were detected on longitudinal magnetization, consequently the fitting model Eq. \ref{amplitude} should be modified to:
\begin{equation}
\label{longitudinal}
\Delta M_\parallel(\theta) \propto \frac{1}{\cos \theta} \exp(-\frac{\pi}{\mu(0) B \cos\theta})
\end{equation}
Here we drop the field-dependent mobility term due to lack of information. Relying on the electronic parameters presented in Ref. \cite{taskin2009quantum}, the carrier mobilities of the 2D surface state ($F_1$ in Fig. \ref{FigBiSb}(a)) and 3D bulk state ($F_2$ in Fig. \ref{FigBiSb}(b)) are 5.5 and 1.2 m$^2$V$^{-1}$s$^{-1}$, respectively. Taking these mobilities as fitting parameters in Eq. \ref{longitudinal}, the curve in Fig. \ref{FigBiSb}(a) can roughly track the fast decrease of dHvA amplitude when field is rotated towards the bisectrix plane down to $\theta \simeq$ 20$^\circ$. However, in Fig. \ref{FigBiSb}(b), the attenuation of dHvA amplitude is obviously much slower than that expected in our 2D model at $\theta <$ 50$^\circ$. It should be mentioned that the frequency $F_2$ comes from a highly anisotropic ellipsoidal FS in Bi$_{1-x}$Sb$_x$: the length of its longest semiaxis (along crystal axis $C_3$)  is 8.5 and 17.7 times of the other two semiaxes, respectively \cite{taskin2009quantum}. The fittings in Fig. \ref{FigBiSb} indicate that even for such an extremely elongated FS, our model can effectively distinguish it from a real 2D cylinder FS.

\begin{figure}[htbp!]
\centering
\includegraphics[width=0.46\textwidth]{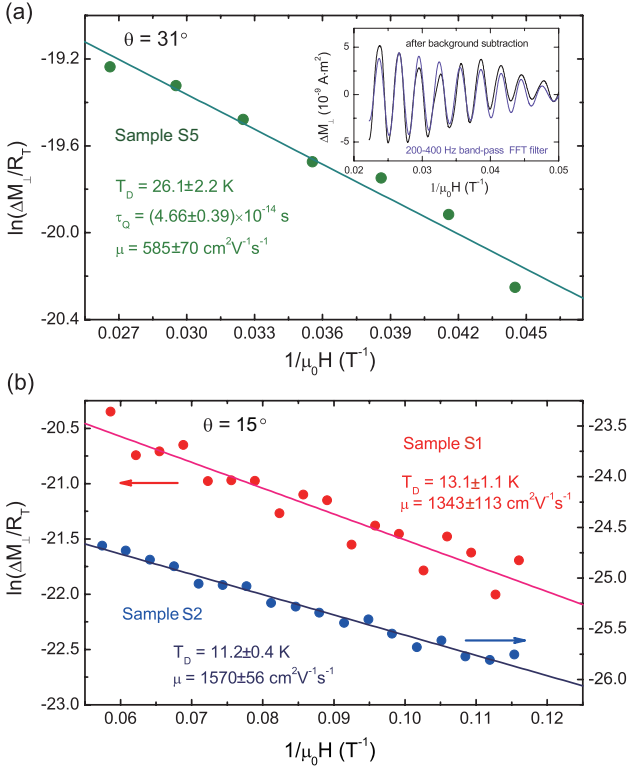}
\caption{The Dingle plots. (a) Dingle plot $\ln(\Delta M_{\perp}/R_T)$ vs. 1/$B$ for sample S5 at $\theta = 31^\circ$. Solid line is the linear fit which has a slope of $-(2\pi^2 k_B m^* T_D)/e\hbar$. The carrier mobility $\mu$ is calculated as $\mu = e\tau_{Q}/m^* = e\hbar/2\pi k_B m^* T_D$. Inset: The raw data of $M_{\perp}$ after subtracting the non-oscillatory background and the band-pass filtered oscillation pattern used in the Dingle plot. (b) Dingle plots for sample S1 and S2 at $\theta = 15^\circ$.
}
\label{FigDingle}
\end{figure}

The 2D LK formula we used in data fitting is theoretically a low magnetic field approximation. In a 2D system, if the number of electrons is kept constant, the chemical potential is prone to be pinned in the highest occupied Landau level and also oscillates with increasing field \cite{shoenberg2009magnetic}. This chemical potential oscillation will cause strong deviation from LK theory at high field, in which the position of chemical potential is assumed to be fixed \cite{harrison1996numerical,champel2001haas}. Such deviation has been found in quasi-2D organic compounds \cite{harrison1996numerical,singleton2000studies,laukhin1995transport} as well as in cuprate high-temperature superconductor YBa$_2$Cu$_3$O$_{6+x}$ \cite{sebastian2011chemical}. In topological insulators, however, this behavior has not ever been reported or discussed. An analytical analogue of LK formula, with well defined $R_T$ and $R_D$, has been established for 2D Dirac-like electron system \cite{sharapov2004magnetic}, and is effectively used in graphene \cite{novoselov2005two}. Conventional LK analysis is widely applied and accepted in the study of quantum oscillations in topological insulators \cite{ando2013topological,taskin2009quantum,analytis2010two,xiong2012high}. As for our data, the upper limit of magnetic field $B$ (45 T) is much lower than the oscillation frequencies $F^\beta$ and $F^\gamma$, therefore the low-field condition $\Delta E_n \ll E_F$ is fulfilled for these two bands (here $\Delta E_n$ is the energy interval between Landau levels and $E_F$ the Fermi energy). Besides, the sharp saw-tooth-like oscillation patterns expect for clean 2D systems are missing in our measurements, and the LK description works well in fitting the temperature dependence of oscillation amplitude \cite{li2014two}. Given the reasonable modeling of the angular dependence of the oscillation amplitude in the surface state of Bi$_{1-x}$Sb$_x$, we conclude that the LK model in our analysis is a correct model.

For the value of $B$ in the fittings, we still use the averaged inverse field \cite{putzke2016inverse} as the effective value: $B_{ave}$ = 17.48 T in Fig. \ref{FigAmpl}(a) and $B_{ave}$ = 6.54 T in Fig. \ref{FigAmpl}(b) and Fig. \ref{FigBiSb}(c). For all three samples, the fittings are reasonably good, though not perfect, with the parameter $\xi$ = 0. We also made curves with finite values of $\xi$ and other fitting parameters unchanged in Fig. \ref{FigAmpl}(a). It appears that an acceptable value is $\xi \lesssim 0.1$, corresponding to $\lambda \lesssim 5.89\times10^{-5}$ and the enhancement on $1/\tau$ is less than $\simeq 12\%$ at 45 T. The small Zeeman effect in scattering rate is qualitatively consistent with the calculation for Bi-based 3D topological insulators \cite{wu2015high}. We note that with an appreciable Zeeman effect there will be visible splitting of the peaks/valleys in the dHvA oscillation patterns. Early work also suggested that a large Zeeman effect makes the Landau Level indexing plot non-linear \cite{taskin2011berry}. None of these effects are observed in our results of the quantum oscillation patterns in SmB$_6$ \cite{li2014two}. Nonetheless, the small mismatch in the fittings in Fig. \ref{FigAmpl} can be assigned to the subtle effects that are ignored in our model such as the Zeeman term, as well as the sample misalignment in the measurement.

The effective fittings by a 2D LK model (Eq. \ref{amplitude}) is an essential evidence against the bulk origin of $F^\beta$. As mentioned above, our model describes a fast amplitude damping with field rotating away from the symmetric axis of FS. The elongated 3D FS in Bi$_{1-x}$Sb$_x$ with FS cross-sectional areas 8.5 times different between two perpendicular directions shows apparent deviation from the fitting curves (Fig. \ref{FigBiSb}(b)). The supposed 3D orbit $\rho$ in FZ SmB$_6$ samples \cite{tan2015unconventional}, which shares the same frequency range with our $\beta$ branch, has a cross-sectional area difference of $\sim$ 3.3 between [101] and [10$\bar{1}$] directions. Such a moderate anisotropy cannot give the fitting results shown in  Fig. \ref{FigAmpl} and Fig. \ref{FigBiSb}(c).

\section{Dingle plot and scattering rate difference between samples}
\label{chap07}

Figure \ref{FigDingle} gives a brief summary of the Dingle temperature $T_D$ in the three samples S5, S1 and S2, derived from the slope of $\ln(\Delta M_{\perp}/R_T)$ versus 1/$B$. The linearity of Dingle plot suggests the field modification on $\tau_Q$ is almost ignorable. While the quantum oscillation mobility $\mu$ = e$\hbar/2\pi k_B m^* T_D$ is 50-100$\%$ larger than the fitting parameter $\mu(0)$ in Fig. \ref{FigAmpl}, the relative magnitude of mobility within the three samples is the same for the two approaches. The discrepancies between mobilities attained by different experimental methods is a famous conundrum in SmB$_6$. In transport measurement, much lower mobilities have been reported, which vary from several tens of cm$^2$V$^{-1}$s$^{-1}$ \cite{luo2015heavy} to 120-140 cm$^2$V$^{-1}$s$^{-1}$ \cite{syers2015tuning,wolgast2015magnetotransport}, and no quantum oscillation has ever been observed. These confusing phenomena may suggest complicated scattering mechanism in this material. Nonetheless, our magnetic quantum oscillation experiment has clearly proved that light carriers with relative high mobilities reside in SmB$_6$, most likely on the surfaces.

The difference of dHvA amplitudes among samples is appreciably large. While sample S5 shows $\Delta M_{eff}$ with a magnitude of $10^{-8}$ A$\cdot$m$^2$ (Fig. \ref{FigTorque}(b)), signals from other samples can be one to two orders of magnitude smaller with the surface area within the same order of magnitude \cite{li2014two}. It is also apparent that the signal strength is not proportional to the related surface area on one sample (Fig. \ref{FigAmpl}(a)). Apart from the most likely reason of surface impurity effect on the carrier mobility, we are also aware of the complex surface reconstruction in SmB$_6$ \cite{ruan2014emergence,rossler2014hybridization,jiao2016additional}. The multiple surface phases are possible to give different contribution to quantum oscillation. This information is not included in the fittings in Fig. \ref{FigAmpl}

\section{Additional information of the temperature dependence of dHvA amplitudes}
\label{chap08}

\begin{figure}[htbp!]
\centering
\includegraphics[width=0.48\textwidth]{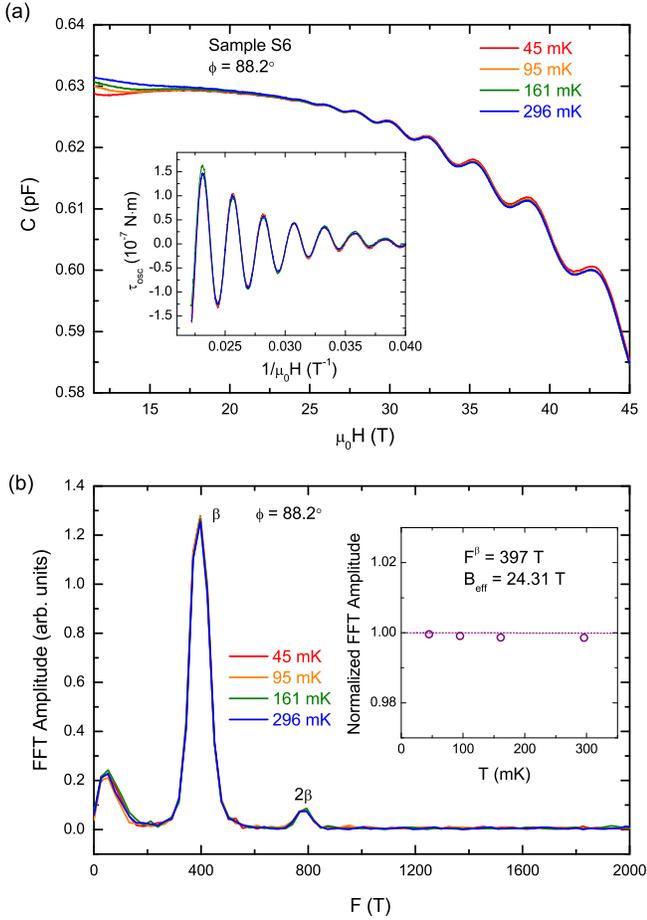}
\caption{(a) Magnetic torque below $^3$He temperature in Sample S6 measured by the cantilever capacitor at $\phi$ = 88.2$^\circ$. Inset: The oscillatory part of magnetic torque plotted against inverse magnetic field. (b) The FFT curves of magnetic torque in a field range between 16.7 T and 45 T. Inset: The normalized FFT amplitudes of $F^{\beta}$ as a function of temperature. Dashed line is fitting based on LK formula with effective mass $m^{*}$ = 0.14 $m_{e}$.
}
\label{FigAddTemp}
\end{figure}

The temperature dependence of the dHvA oscillations in SmB$_6$ is investigated repeatedly in different samples and different magnetic field orientations. In all the measurements the dHvA amplitudes shown almost temperature-independent behavior between the base temperature of the dilution fridge (40-45 mK) and 300 mK. No considerable low-temperature dHvA amplitude increase \cite{tan2015unconventional} has been observed. Figure \ref{FigAddTemp} shows the result taken at $\phi$ = 88.2$^\circ$, i.e., field close to [100] direction, in Sample S6. Similar to the observation shown in Fig. \ref{FigTemp}, the oscillatory magnetic torque curves at all temperatures overlap with each other. The dominating frequency at this tilt angle is $F^{\beta}$ = 397 T as shown in Fig. \ref{FigAddTemp}(b). The amplitude attenuation of this peak is within 0.2$\%$ from 45 mK to 296 mK, in contrast with the previous observation of $>$ 80$\%$ reported by Tan \textit{et. al.} \cite{tan2015unconventional} in floating-zone-grown crystals. At this stage we confirm that such steep increase in dHvA amplitude does not exist in our flux-grown samples.


\end{document}